# A scalable neural bundle map for multiphysics prediction in lithium-ion battery across varying configurations

Zhiwei Zhao[a,b,#], Changqing Liu[a, #], Jie Lin[b], Fan Yang[a], Yifan Zhang[a], Yan Jin[b], Yingguang Li[a, *]

[a] *College of Mechanical and Electrical Engineering, Nanjing University of Aeronautics and Astronautics, Nanjing, 210016, China*

[b] *School of Mechanical and Aerospace Engineering, Queen's University Belfast, Belfast BT9 5AH, UK*

[#] These authors contributed equally to this work.

[*] Corresponding author.

*E-mail address:* liyingguang@nuaa.edu.cn (Yingguang Li).

**Abstract:**

Efficient and accurate prediction of Multiphysics evolution across diverse cell geometries is fundamental to the design, management and safety of lithium-ion batteries. However, existing computational frameworks struggle to capture the coupled electrochemical, thermal, and mechanical dynamics across diverse cell geometries and varying operating conditions. Here, we present a Neural Bundle Map (NBM), a mathematically rigorous framework that reformulates multiphysics evolution as a bundle map over a geometric base manifold. This approach enables the complete decoupling of geometric complexity from underlying physical laws, ensuring strong operator continuity across varying domains. Our framework achieves high-fidelity spatiotemporal predictions with a normalized mean absolute error of less than 1% across varying configurations, while maintaining stability during long-horizon forecasting far beyond the training window and reducing computational costs by two orders of magnitude compared with conventional solvers. Leveraging this capability, we rapidly explored a vast configurational space to identify an optimal battery design that yields a 38% increase in energy density while adhering to thermal safety constraints. Furthermore, the NBM demonstrates remarkable scalability to multi-cell systems through few-shot transfer learning, providing a foundational paradigm for the intelligent design and real-time monitoring of complex energy storage infrastructures.

**Introduction**

Lithium-ion batteries (LIBs) are the cornerstone of the global transition toward sustainable energy systems[1,2]. As the primary energy storage technology for electric vehicles and renewable energy integration, LIBs are subject to rising demands for higher energy density, enhanced safety, and lower costs. Yet, the traditional trial-and-error development cycle remains a significant impediment, as the sheer temporal and financial costs of physical testing cannot sustain the current pace of required innovation[3,4]. The rapid design of next-generation energy storage systems necessitates predictive models that can both reduce experimental overhead and provide deep insights into the complex electrochemical reactions and associated multiphysics processes occurring within the cell[5,6].

Batteries exhibit complicated behaviors during operation arising from the tightly coupled evolution of electrochemical, thermal and mechanical fields, which are highly sensitive to cell geometry, material configuration and boundary conditions[7–9]. Capturing these nonlinear and interdependent processes requires models that can resolve both spatial heterogeneity and temporal dynamics across multiple physical domains[10]. Predictive multiphysics modelling is therefore a valuable tool for guiding battery design, as it enables the evaluation of architectures and operating strategies before costly prototyping and testing[11,12].

Traditional physics-based models, such as the Doyle–Fuller–Newman (DFN) framework, provide high-fidelity insights into these interactions by solving coupled partial differential equations (PDEs)[13]. However, their computational complexity grows exponentially when extended to three-dimensional geometries or complex field interactions, rendering them impractical for real-time monitoring or massive design optimization[14]. To improve efficiency, reduced-order and surrogate models such as the single-particle model (SPM) and its enhanced versions (e.g., SPMe) have been proposed. While reduced-order models offer speed, they often sacrifice accuracy particularly under challenging conditions (e.g., high-rate charging/discharging, large-format cells) where spatial heterogeneity and multiphysics coupling are significant[15,16]. As a result, there remains a gap between physically consistent but slow mechanistic models and computationally efficient but less accurate surrogates[17].

Data-driven approaches have emerged as a promising alternative for accelerating battery modeling by learning complex nonlinear dynamics directly from data. Neural networks have been widely applied to tasks such as state-of-charge estimation, degradation forecasting, and lifetime prediction[18–22]. More recently, physics-informed neural networks have been proposed to embed governing equations into the learning process[23]. Despite these advances, most existing models focus on low-dimensional quantities or fixed geometries, and struggle to accurately represent the spatiotemporal evolution of coupled multiphysics fields across heterogeneous domains[24,25].

Neural operators represent a conceptual shift in scientific machine learning by learning mappings between function spaces rather than between finite-dimensional representations[26]. This operator-level formulation enables generalization across initial conditions and source terms and has shown strong performance in solving PDEs. Neural operators have already been applied to battery prediction, for example using DeepONet[27] and Fourier neural operators[28] to model state-of-charge or concentration fields[29–31]. However, these studies are limited to fixed geometries and individual physical fields, often underpinned by SPMs that cannot capture coupled battery multi-physical fields. Existing variants such as diffeomorphic neural operators have attempted to extend neural operators across different geometries through diffeomorphic mappings to the same reference domain[32,33]. While these methods can theoretically ensure operator continuity within the transformed reference space, they are inherently restricted to geometries that share the same topology and possess differentiable mappings. Another class of methods utilizes Transformer-based architecture to process geometry and physical fields as unified inputs[34,35]. Although these approaches offer greater flexibility in handling diverse geometries, they treat geometry primarily as a discrete feature or field and lack a mathematical framework to guarantee operator continuity across the geometry space. Consequently, both paradigms struggle to overcome the fundamental limitation that functions defined on varying domains belong to distinct Banach spaces, making it theoretically challenging to construct continuous operators across diverse geometries.

In batteries, this limitation is particularly severe. Different physical fields are defined on distinct, geometry-dependent subdomains, such as electrode solid particles, electrolyte and current collectors, and are governed by a mixture of transient and static partial differential equations[36]. As cell geometry or configuration varies, both the topology of these domains and the functional spaces supporting the solution fields change accordingly. As a result, battery evolution is more naturally viewed as a family of geometry-indexed operators acting on heterogeneous state spaces, rather than a single operator defined on a fixed domain. To date, a unified learning framework capable of representing and approximating such geometry-dependent multiphysics evolution remains unavailable.

In this paper, we propose a Neural Bundle Map (NBM) framework and theoretical formulation for multiphysics prediction in LIBs, enabling efficient and accurate learning of the coupled electrochemical–thermal evolution. By generalizing conventional neural operators to a fiber bundle structure, we represent geometric variations on a base manifold and learn multi-physical field mappings within the fiber space. This construction establishes a unified representation across heterogeneous geometries and coupled physical fields. Crucially, NBM internalizes the complex, non-linear coupling of electrochemical-thermal processes by learning the evolution operator itself rather than discrete state mappings. This operator-based approach allows the model to achieve high-fidelity spatiotemporal generalization, where it autonomously forecasts battery states over long-term horizons far exceeding the training window. Mathematically, our framework ensures operator continuity across varying geometries by mapping design parameters onto a compact latent manifold. This continuity guarantees that the physical response remains stable under geometric perturbations, effectively overcoming the Banach space limitations of traditional operators defined on fixed domains. Once initialized with a given state, the neural fiber bundle can autonomously predict subsequent spatiotemporal evolution. When applied to high-energy-density battery design, the NBM identifies optimal configurations within minutes, achieving a 38% increase in energy density while maintaining thermal safety. This framework offers a mathematically rigorous pathway for the intelligent design and monitoring of next-generation battery systems.

## Results

**Framework overview and flowchart**

We developed an NBM framework to predict the coupled multiphysics evolution of LIBs across varying geometries. Given the battery geometry and the multiphysics state at the initial time, the learned NBM autonomously generates the subsequent spatiotemporal evolution of all coupled fields (Fig. 1a). The conceptual foundation of the NBM is rooted in the fiber bundle theory, where the battery design space is represented as a base manifold $\mathcal{B}$. Each point $b \in \mathcal{B}$ denotes a specific battery geometric configuration, such as cell length, cell width, and electrode thickness, which determines the fiber $\mathcal{F}_b$. This fiber $\mathcal{F}_b$ represents the total space of coupled electrochemical and thermal fields in the cell evolving over time. The evolution of the battery state is then modeled as a bundle map $\Psi: \mathcal{E} \to \mathcal{E}$ that propagates the multi-physical states along fibers while maintaining geometric consistency (Fig. 1b).

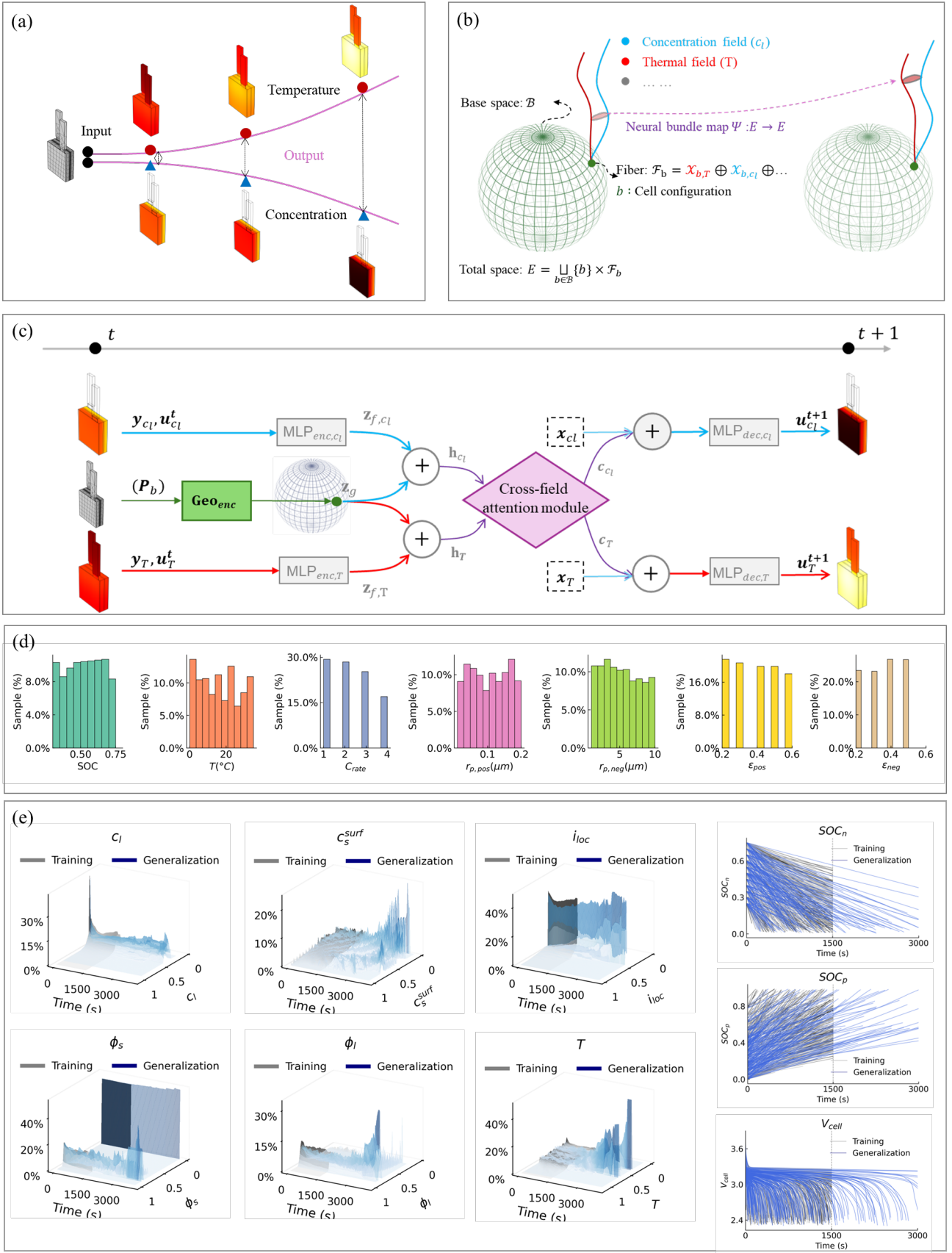

Fig. 1 | The flowchart of the proposed NBM for lithium-ion battery multiphysics filed estimation. (a) Multiphysics rollout and forecasting. Once initialized, the model autonomously predicts the spatiotemporal evolution of potential fields without intermediate

numerical solvers. (b) Conceptual foundation based on fiber bundle theory. The battery design space is represented as a base manifold $\mathcal{B}$, where each point $b$ denotes a unique geometric configuration. The associated fiber $\mathcal{F}_b$ encapsulates the total space of coupled electrochemical (e.g., electrolyte concentration $c_l$) and thermal fields ($T$). The NBM $\Psi: \mathcal{E} \to \mathcal{E}$ acts as a bundle map that propagates multiphysics states forward in time while maintaining geometric consistency across different cell architectures. (c) Architecture of NBM. The framework integrates a geometry encoder (e o$_{enc}$) that maps high-dimensional cell geometries into a compact latent manifold $z_g$ on a unit hypersphere, ensuring operator continuity under geometric perturbations. Modular field encoders (MLP$_{enc}$) transform spatial physical states into latent features $z_f$, which are then fused with $z_g$ to form geometry-conditioned representations $h$. A cross-field attention module approximates the geometry-dependent kernel functions to resolve the non-linear coupling between disparate physical domains. The updated states $u_{t+1}$ are reconstructed via field-specific decoders (MLP$_{enc}$).

The architecture of the NBM is designed to mirror the physical coupling of the underlying governing equations through modular neural components (Fig. 1c). The construction of the NBM is governed by strict theoretical requirements regarding geometric regularity and integral kernel approximation. To ensure that the physical evolution varies continuously with respect to geometric perturbations, a geometry encoder module ($Geo_{enc}$) is used to map the battery design space onto a compact latent manifold $z_g \in \mathcal{B}$, specifically a unit hypersphere, which serves as a locally compact Hausdorff base space. This embedding must be Lipschitz-admissible to preserve the strong operator continuity of the physical fields across different cell configurations. Within this structure, the governing partial differential equations are reformulated into a unified Fredholm integral framework. Independent sub-field encoders (MLP$_{enc}$) learn representations for physical variables, which are then fused with $z_g$ using concentrate mechanism. Then a cross-field attention module acts as a neural approximation of the geometry-dependent kernel function $K(x, y; b)$, while a cross-field attention module approximates the block-kernel matrix representing inter-domain couplings (Fig. 1b). This allows both transient fields, such as the thermal field, and static fields, such as the electrolyte potential, to be resolved through a canonical kernel-based operator family. Finally, a specific point-query based decoder (MLP$_{dec}$) is used to predict the corresponding fields.

Consequently, NBM operates as an operator family-based predictor. This operator formulation allows the model to propagate system dynamics forward in time without repeated numerical solving. Detailed information can be found in the Method section.

**Data generation**

To generate a comprehensive dataset for training and evaluating the NBM, we consider a representative large-format 20Ah lithium-ion prismatic pouch cell (A123 System) composed of a LiFePO$_4$ positive electrode, and a graphite negative LiPF$_6$ electrode. Specifically, the training data were generated using a computationally efficient high-fidelity 3D physics-based model[36]. This model streamlines the classic DFN framework through a dimensional analysis while incorporating cell multiscale coupling of electrochemical and thermal processes. By resolving the interdependent dynamics of ion transport, charge transfer, and heat generation, the model provides high-accuracy ground truth data. All simulations are performed under constant-current discharge, starting from an initial terminal voltage of 3.7 V and terminating when the voltage reaches 2.5 V. This setup provides a physically consistent foundation for capturing the interdependent dynamics of the electrochemical and thermal states under realistic operating conditions.

The geometric configuration of the battery is systematically varied to account for structural heterogeneity relevant to practical battery design, including variations in cell dimensions such as width $W_{cell}$ and height $H_{cell}$, as well as the thicknesses of current collectors and the specific geometries of tabs and busbars, as summarized in Table S 1. These geometric variations are selected to represent different battery form factors and design choices commonly encountered in cell-level and module-level configurations, which are known to influence current distribution, voltage response, heat generation, and heat dissipation. In addition to geometric parameters, operating and initial conditions are sampled across a broad range, including initial temperature and discharge C-rate. All parameters are generated through uniform random sampling within predefined physical bounds, and their distributions are shown in Fig. 1d, ensuring sufficient coverage of both typical and extreme operating regimes.

For each simulated case, the spatiotemporal evolution of multiple physical fields is recorded throughout the discharge process. The output variables include temperature $T$, solid-phase electric potential $\phi_s$, electrolyte potential $\phi_l$, local current density $i_{loc}$, average solid-phase lithium concentration at the particle surface $c_s^{surf}$ and electrolyte concentration $c_l$, together with boundary-level and integral quantities such as cell terminal voltage $V_{cell}$ and the state of charge of both positive $SOC_p$ and negative $SOC_n$ electrodes. Because different physical fields are defined over distinct spatial regions, all variables are extracted in a domain-aware manner, with field quantities recorded on their corresponding physical subdomains and global quantities tracked at the cell or electrode level.

All simulations are sampled at a fixed temporal resolution of 10 s. To evaluate the robustness and temporal extrapolative capacity of the framework, we established a distinct separation between the training and testing data horizons. The training dataset comprises 1,000 simulated cases with a truncated discharge duration of 1,500 s, focusing on the initial and mid-stages of electrochemical evolution. In contrast, the test set consists of 200 cases extending to a maximum duration of 3,000 s, effectively doubling the temporal window to assess long-horizon stability.

To quantify the challenge of this generalization task, we first analyze the data distribution differences between the training and test sets as shown in Fig. 1e. A noticeable disparity in data distributions is observed between the 1,500 s training window and the 1,500 to 3,000 s generalization window. To quantify this difference, we calculate the histogram overlap between distributions. For the solid-phase potential $\phi_s$, the electrolyte potential $\phi_l$, and local current density $i_{loc}$, the overlap values are below 0.5, as illustrated in Table S 2. This indicates that the model is required to generalize across data distributions that are significantly different from those encountered during training.

**Geometric manifold structure in the latent base space.**

We analyze the structure of the learned geometric latent variables to determine whether the NBM framework captures a physically meaningful representation of the battery configuration space. In the NBM architecture, battery geometries are embedded into a latent manifold defined on a hypersphere, where the distances along the manifold reflect the underlying similarity between different geometric configurations. A fundamental prerequisite for operator continuity is that similar geometric configurations in the physical design space must remain proximal within the learned latent manifold to ensure the stability of the predicted multiphysics evolution.

To validate this property, we conduct controlled variations of the battery geometry by systematically perturbing specific design parameters while keeping others constant. For each configuration, the corresponding geometric latent variables are extracted and pairwise distances on the hyper-spherical manifold are computed using geodesic distance. These manifold distances are then compared against the normalized Euclidean distances of the corresponding geometric parameters in the original design space.

The relationship between the physical geometric distance and the manifold geodesic distance is presented in Fig. 2a for three representative groups of geometric variations: cell dimensions, tab geometry, and busbar geometry. In all instances, a clear positive correlation exists between the physical design space and the learned manifold, which confirms that the NBM accurately preserves the topological relationships of the input geometries. Specifically, variations in cell geometry, the correlation coefficient reaches 0.83, which indicates a robust alignment between geometric similarity and manifold proximity. Variations in busbar geometry exhibit a moderate correlation of 0.66, while tab geometry variations yield a correlation of 0.71.

Beyond these quantitative correlations, the learned manifold captures the relative sensitivity of multiphysics behavior to different geometric components. The average geodesic distance induced by variations in cell geometry exceeds those associated with bar and tab geometries. This ordering is consistent with the physical hierarchy of geometric influence on electrochemical and thermal fields, where global cell dimensions exert the strongest effect, followed by current collectors and then tabs. The structural consistency between the physical sensitivity and the latent manifold distance demonstrates that the NBM encodes geometry in a rigorous and physically intuitive manner. By preserving neighborhood relationships between battery geometries, the learned representation supports smooth variation of the operator with respect to geometry, providing a foundation for robust generalization across design spaces and ensuring continuity of multiphysics state evolution with respect to geometric perturbation.

**Spatiotemporal prediction fidelity and extrapolative generalization**

We evaluate the spatiotemporal prediction fidelity of the NBM by examining its performance across the full battery domain and extended time horizons for all test cases.

Accurate spatial prediction is particularly critical for battery analysis, such as non-uniform heat generation can lead to the formation of localized hot spots, which accelerate degradation and may trigger thermal runaway if not properly identified. Consequently, a robust predictive model must resolve spatially localized features and gradients across the battery domain. To provide a rigorous pointwise assessment, we analyzed a random subset of 100,000 spatiotemporal sample points drawn from the test set, spanning all spatial regions and time steps. As illustrated in Fig. 2b, the predicted values for most physical fields align closely with the reference solutions. Although moderate dispersion is observed in certain electrochemical quantities, such as the electrolyte potential $\phi_l$ and terminal voltage $V_{cell}$, the error spread remains bounded without a systematic bias toward overprediction or underprediction. The overall error spread remains bounded across the entire space-time domain. Quantitatively, the average NMAE across the test set is approximately 0.2% for solid-phase potential $\phi_s$, solid-phase surface concentration $c_s^{surf}$ and local current density $i_{loc}$, as shown in Table 1. The errors are slightly higher for other fields including the temperature $T$ (0.5%), $\phi_l$ (0.7%) and $V_{cell}$ (0.9%).

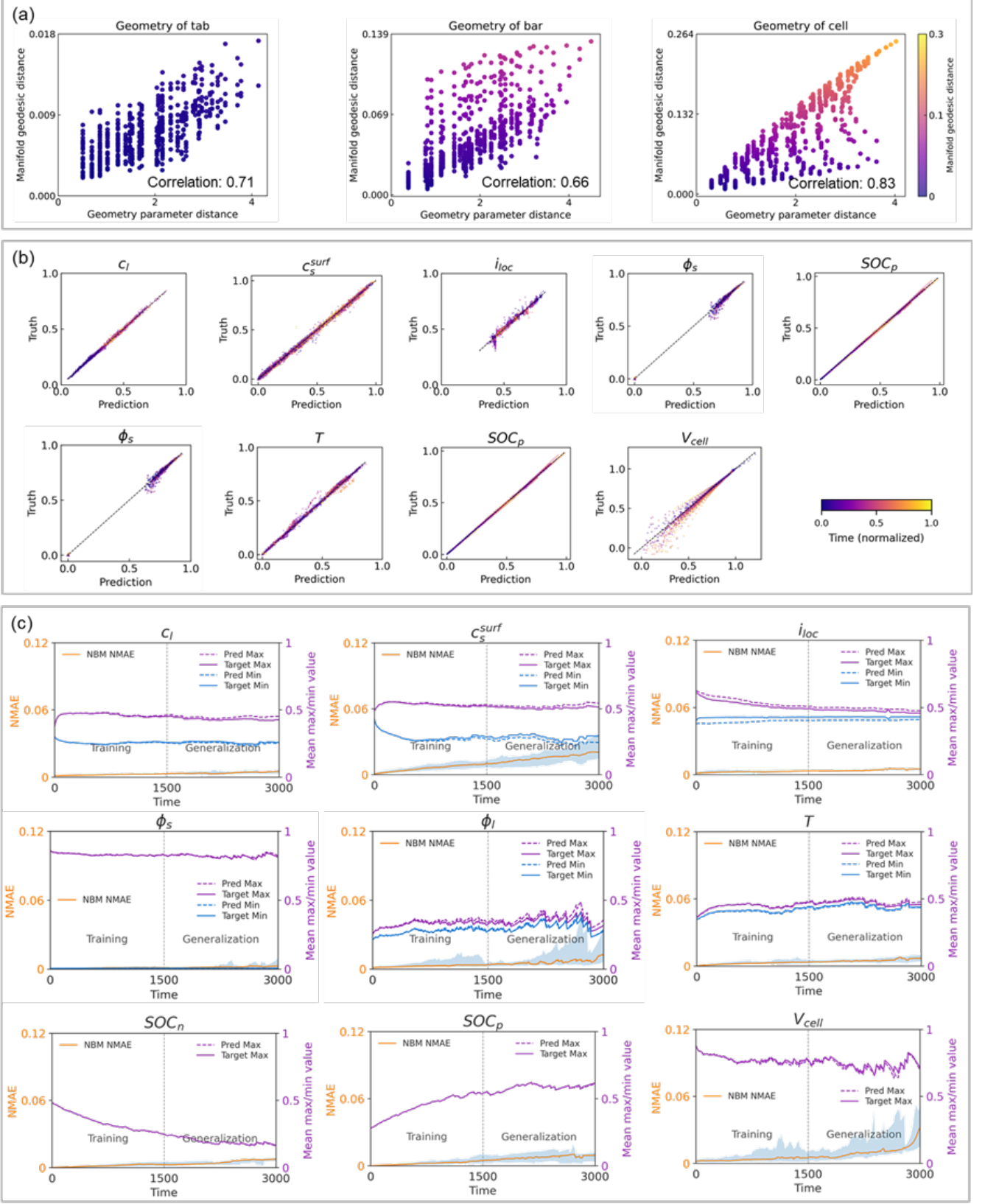

Fig. 2 | Spatiotemporal fidelity and extrapolative robustness of the NBM framework. (a) Topological preservation in the latent base

space. Pairwise geodesic distances on the learned manifold are compared against normalized physical design distances for tab, busbar, and cell geometries. Clear positive correlations (up to 0.83 for cell geometry) demonstrate that the NBM accurately internalizes the topological relationships of the design space, ensuring operator continuity under geometric perturbations. (b) Pointwise assessment of multiphysics prediction accuracy. The scatter plots compare NBM predictions against reference numerical solutions across a random subset of 100,000 spatiotemporal sample points from the test set. (c) (c) Long-horizon temporal generalization and stability. The model, trained on the initial 1,500 s of discharge, is tested on an extended 3,000 s window to evaluate extrapolative capacity. It shows the evolution of temporal average NMAE and the calibration of predicted maximum values against ground truth.

The ability to maintain such fidelity over extended periods is fundamental for capturing multiphysics dynamics. For effective design and optimization, it is essential that the learned operator generalizes well beyond the training time window, maintaining accuracy even as the system undergoes prolonged evolution. To assess the extrapolative robustness of the NBM, the model was trained on data from the first 1,500 s of discharge and tested on a 3,000 s horizon, effectively doubling the temporal window. We evaluate the temporal prediction accuracy using the average NMAE and the evolution of the maximum values of key physical fields.

Despite data distribution shifts (as shown in Data generation Section), the NBM retains high prediction accuracy as the simulation progresses into the unseen temporal regime. As illustrated in Fig. 2c, the average NMAE for fields such as the electrolyte concentration $c_l$, the local current density $i_{loc}$, and the temperature $T$ remains low throughout the entire 3,000 s duration. While the errors for the terminal voltage $V_{cell}$, $\phi_l$ and $c_s^{surf}$ increase slightly in the late stages of discharge, they remain below 0.06.

Furthermore, the predicted maximum and minimum normalized values are well calibrated against the ground truth, particularly for fields with significant distribution differences such as $V_{cell}$, $\phi_l$, $\phi_s$ and $T$. These results indicate that the model effectively captures the underlying temporal dynamics and successfully extends those dynamics to the later stages of discharge. The consistency of prediction errors over time ensures that the trajectories remain stable and do not diverge, highlighting the reliability of the learned bundle map for long-term forecasting and its ability to generalize to unseen temporal domains.

Overall, these results confirm that the model has learned the temporal dynamics of the battery and successfully extended those dynamics to later stages of discharge. The consistency of prediction errors over time ensures that the predictions remain stable and do not diverge, highlighting the reliability and robustness of the learned operator for long-term forecasting. This indicates that the neural bundle map effectively generalizes unseen time periods, maintaining high prediction accuracy across an extended temporal range.

Table 1 Test set performance metrics.

| Physics fields | $T$ | $c_l$ | $\phi_l$ | $c_s^{surf}$ | $V_{cell}$ |
|---|---|---|---|---|---|
| Average NMAE | 0.5% ± 1.2% | 0.2% ± 0.3% | 0.7% ± 1.8% | 0.1% ± 0.2% | 0.9% ± 2.3% |
| Physics fields | $\phi_s$ | $SOC_p$ | $SOC_n$ | $i_{loc}$ | |
| Average NMAE | 0.2% ± 0.5% | 0.2% ± 0.3% | 0.2% ± 0.2% | 0.2% ± 0.3% | |

**Validation of Multiphysics evolution against experimental measurements and simulation results.**

We evaluate the ability of the proposed NBM to predict the full multiphysics state evolution of a real battery cell during a complete discharge cycle. In this context, the specific battery geometry and the complete multiphysics state at the initial time are provided as inputs, after which the model autonomously generates the subsequent evolution of all coupled physical fields through a recursive application of the learned NBM. Notably, no exogenous temporal information or intermediate physical states are provided during the inference phase, rendering this task a rigorous long-horizon rollout of the underlying battery dynamics. To rigorously assess fidelity of the model in a physical environment, the predictions are compared not only against the numerical simulations (Truth) but also against experimental measurements of surface temperature and terminal voltage.

Fig. 3a illustrates the spatiotemporal temperature evolution on the battery surface at two critical time steps ($t$=400 s and $t$=800 s). The NBM accurately reproduces the global temperature rise and the non-uniform spatial distribution observed in experimental measurements. Specifically, the model resolves the localized heating near the tabs and the resulting thermal gradients of the cell. Quantitatively, the NBM achieves exceptional thermal prediction accuracy, at a discharge end state ($t$=800 s), the spatially averaged NMAE relative to experimental measurements is 0.77%, corresponding to a degree mean absolute error (MAE) of 0.39 °C. Notably, this performance closely approaches the intrinsic error of the high-fidelity numerical simulation relative to the experiment (0.49%), while the NMAE between the NBM prediction and simulation truth is minimized at 0.50%.

Furthermore, the NBM demonstrates high precision in predicting the temporal trajectory of the terminal voltage $V_{cell}$, as shown in Fig. 3c. Throughout the entire discharge duration, the model yields a temporal average NMAE of 1.73% relative to experimental measurements. As the system approaches the discharge cutoff, the NBM exhibits alignment with experimental data, achieving a terminal voltage error of only 0.0015 V. These results demonstrate that the NBM not only preserves the rigorous computational fidelity of traditional solvers but also maintains high consistency with physical experimental benchmarks throughout the discharge process.

Regarding the internal electrochemical fields where direct experimental measurement remains a challenge, such as electrolyte concentration $c_l$ and solid-phase surface concentration $c_s^{surf}$, the NBM maintains strict consistency with the simulation truth. As depicted in Fig. 3b, the model successfully captures the accumulation dynamics and spatial contrast within the positive and negative electrodes. For these region-specific fields, the spatially averaged NMAE remains below 1.1% throughout the discharge process. Similarly, the prediction of electrode-level states, $SOC_p$ and $SOC_n$, exhibits stable and continuous evolution without numerical drift. Quantitatively, the prediction for $SOC_p$ and $SOC_n$ yields time-averaged NMAEs of 0.3% and 0.8%, respectively. These errors remain bounded even as the system approaches the discharge cutoff, indicating that the learned NBM effectively preserves the long-term temporal coherence of the coupled multiphysics system.

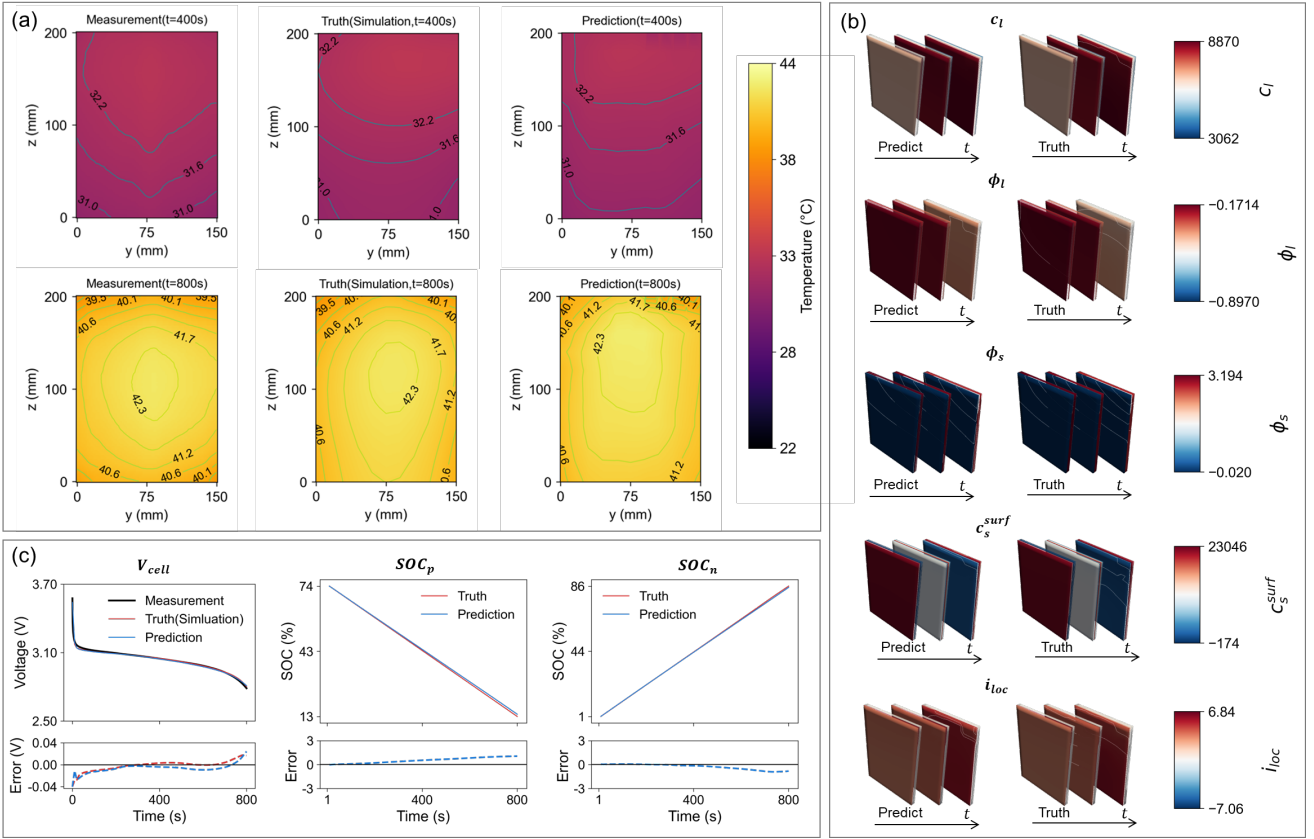

Fig. 3 | Battery state evolution prediction. (a) Spatiotemporal evolution of the battery surface temperature field. The figures compare experimental thermal measurements (left), numerical simulations (center, Truth), and NBM predictions (right) at two critical timestamps during discharge: $t = 400$ s and $t = 800$ s. The NBM accurately reproduces the global temperature rise and localized heating gradients. At the end of discharge. (b) Three-dimensional reconstruction of internal coupled multiphysics fields. The sequences show the autonomous rollout of the electrolyte concentration $c_l$, electrolyte potential $\phi_l$, solid-phase potential $\phi_s$, solid-phase surface concentration $c_s^{surf}$, and local current density $i_{loc}$. (c) Temporal trajectories and error analysis of global battery states. Top panels show the evolution of terminal voltage $V_{cell}$, positive electrode state-of-charge $SOC_p$, and negative electrode state-of-charge $SOC_n$.

**Battery design optimization with NBM**

In this section, we evaluate the practical utility of the Neural Bundle Map (NBM) framework in high-dimensional battery design optimization. The primary objective is to solve a multi-objective optimization problem that aims to maximize gravimetric energy density $E_d$ while simultaneously minimizing the maximum temperature $T_{max}$ during the discharge process. This optimization task involves varying a diverse set of design parameters including electrode geometries, material properties, and initial operating conditions.

We begin by analyzing the sensitivity of $E_d$ and $T_{max}$ to individual geometric parameters and configurational parameters. Fig. 4a-c illustrate the relationships between specific parameters, such as electrode geometry ($H_{cell}$ and $W_{cell}$), particle radii ($r_{p,pos}, r_{p,neg}$), and electrode porosities ($\epsilon_{pos}, \epsilon_{neg}$), and the resulting battery performance. The results demonstrate that global cell dimensions and electrode porosities exert the most significant influence on the energy-thermal trade-off.

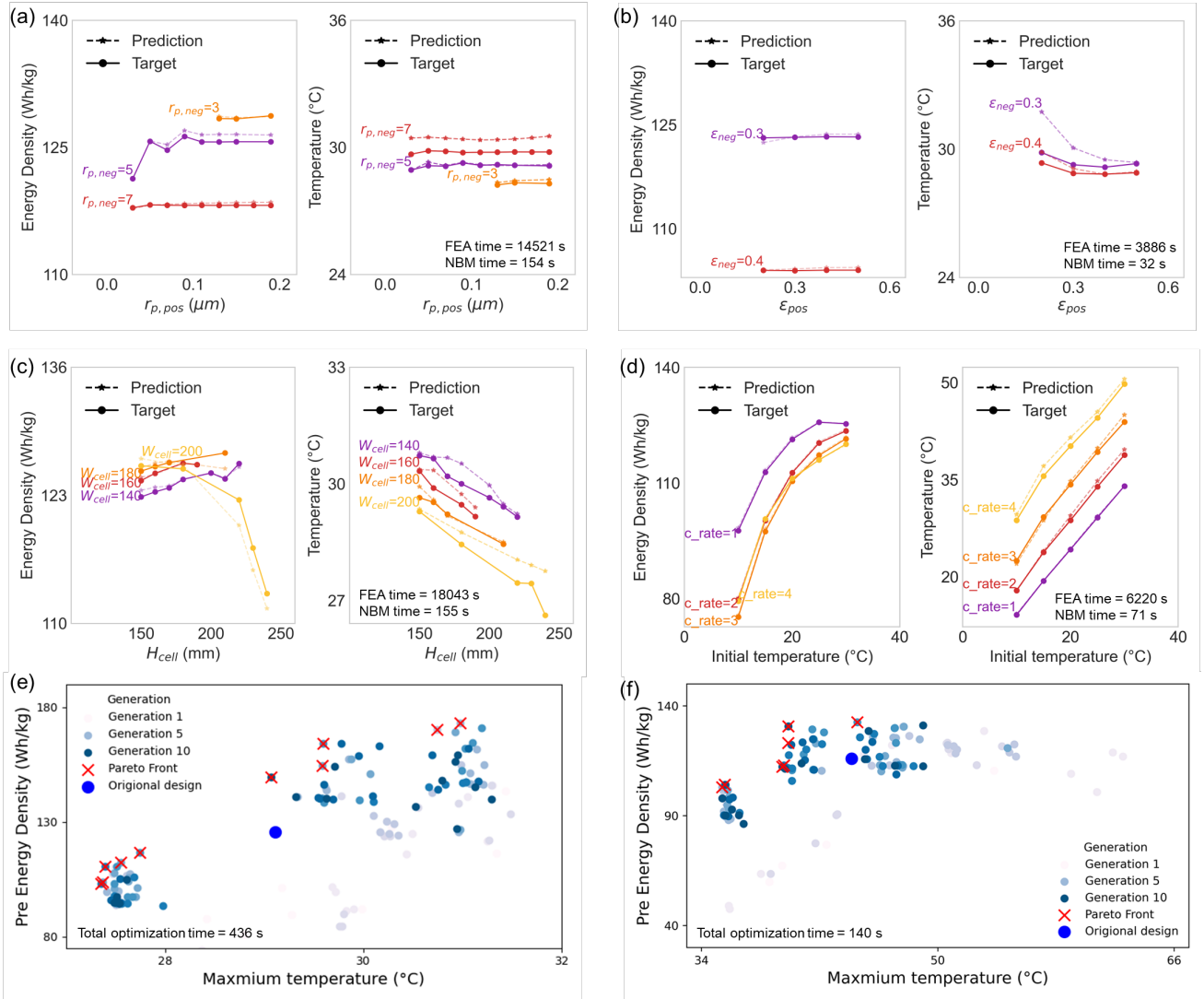

Fig. 4 | Multi-objective design optimization and sensitivity analysis enabled by the NBM. (a–c) Sensitivity analysis of gravimetric energy density $E_d$ and maximum temperature $T_{max}$ relative to electrode microstructural parameters (particle radii $r_{p,pos}, r_{p,neg}$), porosities ($\epsilon_{pos}, \epsilon_{neg}$), and cell geometries ($H_{cell}$ and $W_{cell}$). The NBM predictions (dashed lines) show high alignment with reference FEA simulations (solid lines) across the design space. (d) Impact of operational variables, including discharge C-rates and initial temperatures, on battery performance metrics. (e, f) Multi-objective optimization results using the NSGA-2 algorithm for 1C (e) and 4C (f) discharge scenarios. The NBM-driven search successfully identifies Pareto-optimal configurations (red crosses) that significantly surpass the baseline design (blue circle). Specifically, at a 1C rate, the framework discovers a configuration achieving a 38% increase in $E_d$ with only a marginal 6% rise in $T_{max}$.

The NBM predictions align closely with the reference physical simulations across the entire design space. Missing data points in the figures are due to convergence issues in the simulation process. Critically, the NBM performs these evaluations with a computational speed that is two orders of magnitude faster than traditional finite element analysis (FEA). This reduction in latency enables a rapid and granular exploration of the configurational manifold that would be computationally prohibitive using conventional solvers.

Furthermore, we also examine the effect of discharge rate and initial temperature on battery performance, as illustrated in Fig. 4d. The model shows consistent prediction accuracy under varying discharge conditions, confirming its robustness when exposed to different operating environments. For example, increasing the initial temperature leads to higher energy density but also raises the maximum temperature, emphasizing the trade-off between energy density and temperature control.

For the multi-objective optimization, we use the NSGA-2 optimization algorithm with 10 generations and 15 populations to search for optimal solutions. The optimization is performed under two different discharge rates: 1C and 4C, and the resulting Pareto fronts are shown in Fig. 4e and Fig. 4f.

For the 1C discharge case, the NBM successfully identifies an optimized configuration that achieves a 38% increase in $E_d$ compared to the baseline design. This improvement is attained with a marginal increase in $T_{max}$ of only 6%. In the high-rate 4C discharge scenario, the model identifies a design that enhances $E_d$ by 14% while maintaining a minimal temperature rise of just 1%. These results are summarized in Table 4 and Table 5, which detail the specific geometric parameters and performance metrics for the selected optimal solutions.

The entire multi-objective optimization process, involving hundreds of iterations across the design space, is completed within several minutes. The ability of the NBM to generalize across varying C-rates and initial temperatures ensures that the optimized designs remain robust under diverse operational environments. These findings demonstrate that the NBM provides a powerful and mathematically rigorous tool for the accelerated design of next-generation energy storage systems where performance and safety must be balanced.

**Extension to multi-cell battery pack systems.**

To further evaluate the architectural flexibility and scalability of the NBM framework, we extend the trained single-cell operator to predict the multiphysics evolution of complex battery packs. We consider parallel-connected configurations comprising 3 to 10 cells, where each cell inherits the geometric and material variability of the original design space.

A key advantage of the NBM is its ability to maintain a consistent network structure while adapting to system-level interactions. From a physical perspective, electrochemical processes, such as concentration gradients and potential distributions, remain predominantly coupled within individual cells. Conversely, the thermal field exhibits global coupling across the pack due to heat conduction and shared boundary conditions between adjacent cells. To reflect this hierarchy, we adapt the cross-field attention module: intra-cell electrochemical fields follow the single-cell coupling logic, while the thermal field utilizes a global attention mechanism to incorporate the temperature states of all cells within the pack.

The transferability of the pre-trained NBM is validated through a few-shot fine-tuning strategy. As illustrated in Fig. 5a, we compare the performance of models fine-tuned with 50, 100, 200, and 300 samples against a baseline model re-trained from scratch using 300 samples. Remarkably, the fine-tuned NBM with as few as 50 samples achieves significantly lower NMAE across all physical fields compared to the fully re-trained model. This suggests that the NBM has successfully internalized the fundamental governing laws of battery physics, requiring only minimal data to adapt to the boundary conditions of multi-cell systems

Quantitatively, with 300 fine-tuning samples, the temporal average NMAE for critical fields such as temperature $T$ and state-of-charge remains below 0.5%. We further analyze the spatial error distribution across a 10-cell pack (Fig. 5b). By scaling the battery index to a normalized [0, 1] interval, the NBM demonstrates uniform prediction accuracy across the entire system. There is no observable error accumulation as the number of cells increases, confirming that the fiber bundle formulation effectively decouples individual cell complexity from global system dynamics. These results position the NBM as a powerful foundation for the monitoring and safety management of large-scale energy storage systems.

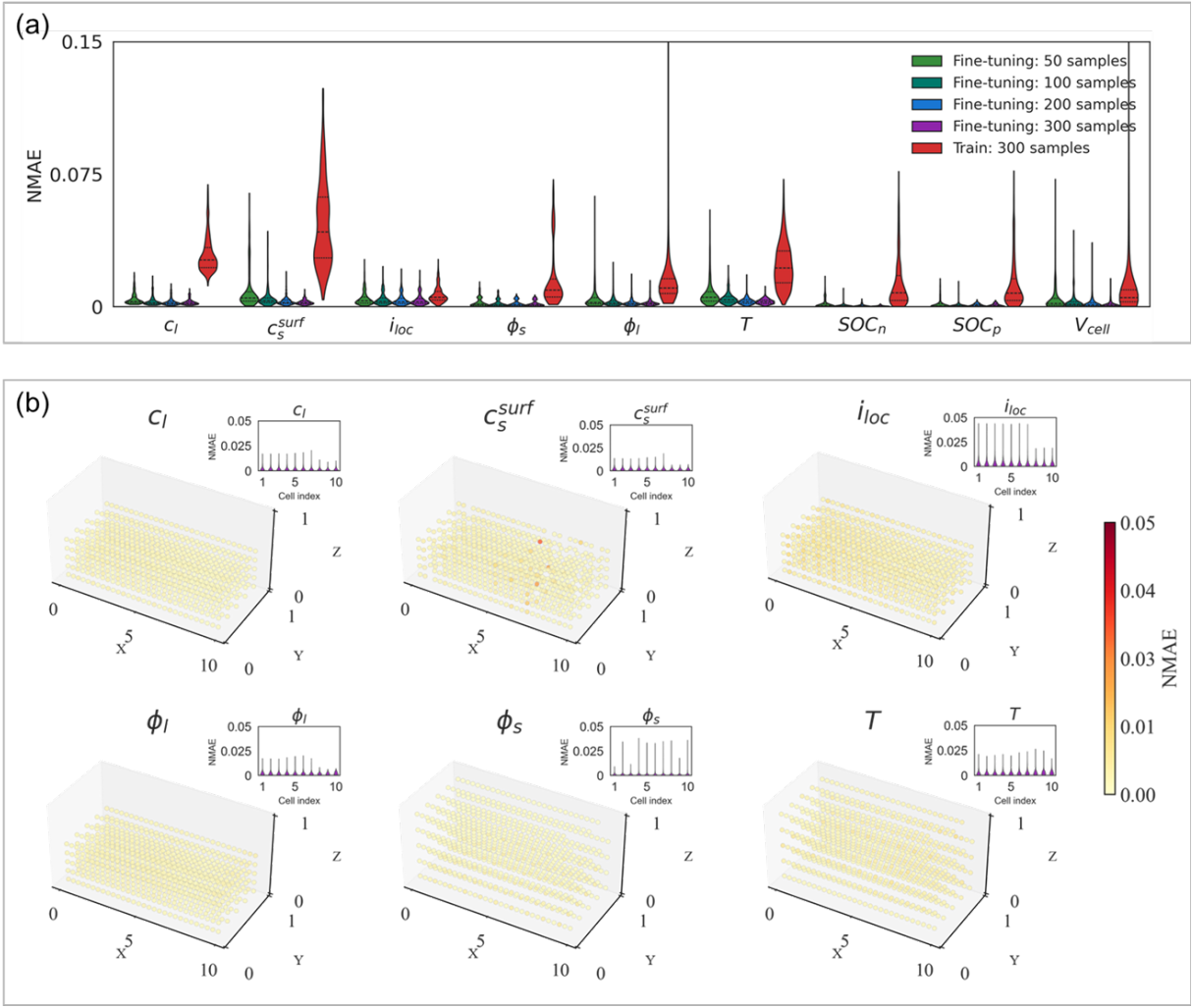

Fig. 5 | Scalability and few-shot transfer learning to multi-cell battery packs. (a) It illustrates the NMAE for physical fields in multi-cell systems. The fine-tuned NBM, using as few as 50 samples, significantly outperforms a model re-trained from scratch with 300 samples, indicating that the framework has internalized the fundamental governing laws of battery physics. (b) Spatial prediction fidelity in battery packs. The uniform NMAE across individual cells (normalized index [0, 1]) demonstrates that the NBM effectively decouples cell-level complexity from global system-level thermal and electrochemical interactions.

**Limitations and outlook**

Although the current investigation demonstrates the high fidelity of the NBM framework for LiFePO$_4$/graphite battery systems, the methodology remains inherently adaptable to diverse battery chemistries. Future research may extend this operator-based approach to Nickel Manganese Cobalt (NCM) cathodes or all-solid-state architectures where different interfacial physics dominate. The inherent flexibility and scalability of the NBM allow for the rapid evaluation of varied material configurations and their subsequent impacts on gravimetric energy density and thermal management.

While this study utilized a specific set of design parameters, the underlying mathematical formulation is applicable to a significantly broader design space. For example, additional parameters, such as tab geometry and electrode designs, or even factors in battery pack design, can be incorporated into the optimization process. The neural bundle map can be applied to these more complex scenarios, offering a pathway for optimizing multi-parameter designs and accelerating the development of next-generation battery systems.

Furthermore, the framework could be adapted to investigate critical practical challenges in the battery industry. Potential applications include the optimization of extreme fast-charging protocols, the prediction of long-term capacity degradation over thousands of cycles, and the integration of advanced functional materials. The generalizability of the NBM across heterogeneous configurations and operational requirements positions it as a robust tool for both fundamental academic research and commercial battery engineering.

**Discussion**

The ability to rapidly predict the spatiotemporal evolution of coupled multiphysics fields in LIBs is essential for accelerating design cycles and enabling real-time performance monitoring. Traditional methods often require time-consuming simulations and experimental testing, which makes real-time optimization of battery systems difficult. In this work, we propose a NBM framework based on fiber bundle theory that unifies electrochemical and thermal dynamics within a rigorous mathematical structure, enabling fast design iterations and optimization.

The NBM framework represents a significant advancement in scientific operator learning by addressing the fundamental limitations of traditional neural operators. While existing models are often restricted to fixed domains or single physical fields, the NBM utilizes a unified Fredholm integral representation to resolve the interdependence of transient and static partial differential equations. Our theoretical analysis confirms that this architecture preserves strong operator continuity across varying geometries, ensuring that the physical response remains stable under geometric perturbations. This rigorous foundation allows the model to achieve a temporal averaged normalized mean absolute error (NMAE) of less than 1% for critical fields.

From an engineering perspective, the NBM facilitates a paradigm shift in battery design by reducing computational latency by two orders of magnitude compared to conventional finite element solvers. This acceleration enables complex multi-objective optimization tasks that were previously impractical. Our case study on high-energy-density battery design identifies configurations that increase gravimetric energy density by 38% while maintaining strict thermal safety constraints. Furthermore, the model demonstrates remarkable robustness in temporal generalization, accurately forecasting battery states over long-term horizons that significantly exceed the training window.

In summary, the Neural Bundle Map provides a scalable and mathematically principled pathway for the design and monitoring of next-generation energy storage systems. By bridging the gap between high-fidelity mechanistic modeling and rapid data-driven inference, this framework offers a powerful tool for optimizing performance, safety, and longevity across various battery chemistries and form factors.

## Methods

Data-driven approaches for solving partial differential equations (PDEs) have demonstrated substantial success in systems constrained to fixed spatial domains or single-field evolutions. However, complex engineering systems such as electrochemical cells present a fundamental challenge to these conventional frameworks. LIBs are characterized by the concurrent evolution of heterogeneous physical fields, including the thermal field $T$, solid-phase potential $\phi_s$, electrolyte potential $\phi_l$, and concentration fields $c_l$. These variables are defined across distinct, geometry-dependent subdomains such as porous electrode continua and solid particles. Furthermore, the governing laws comprise a mixture of transient equations, such as heat and diffusion, and static equations, such as elliptic potential models.

This combination of heterogeneous fields, disjoint domains, and varying geometries renders existing operator learning frameworks insufficient. Models trained on a fixed domain typically fail to generalize to new geometrical configurations. Additionally, methods that require all fields to be defined on a single shared domain cannot fully capture the underlying physical coupling structure of the system. These limitations underscore a fundamental necessity for a unified mathematical and computational formalism capable of representing the evolution of multi-field and multi-domain systems subject to geometric variability.

To address this, a fiber-bundle formulation for multiphysics evolution is introduced, wherein dynamics are learned as a bundle map via neural approximation. This formulation provides a mathematically principled foundation for representing heterogeneous physical fields on geometry-dependent partitions, expressing their joint evolution under coupled PDEs, and enabling the generalization of data-driven models across diverse cell architectures.

## Fiber bundle formulation of geometry-dependent multiphysics

The mathematical modeling of electrochemical batteries is fundamentally characterized by the dependence of heterogeneous physical fields on the specific geometric configuration of the cell. To rigorously encode this geometric variability within a unified framework, the collection of all admissible multiphysics state spaces is formalized as a Hilbert fiber bundle over a geometry base space.

Let $\mathcal{B} \subset \mathbb{R}^d$ represent the manifold of geometric configuration parameters, where each $b \in \mathcal{B}$ parameterizes a specific cell architecture, such as electrode thickness or particle radius. This parameterization determines the topological and metric properties of the underlying computational subdomains. To account for multiphysics heterogeneity, let $\mathcal{A}$ represent the index set of physical fields (e.g., solid-phase concentration, electrolyte potential, and temperature). For a fixed geometry $b \in \mathcal{B}$ and a specific physical field $a \in \mathcal{A}$, the field variable is defined on a configuration-dependent subdomain $\Omega_{b,a} \subset \mathbb{R}^3$.

Associated with each domain is a Hilbert space $\mathcal{X}_{b,a}$. The full multiphysics state space for a specific configuration $b$, denoted as the fiber $\mathcal{F}_\mathbf{b}$, is constructed as the direct sum of these heterogeneous function spaces,

$$\mathcal{F}_\mathbf{b} = \bigoplus_{a \in \mathcal{A}} \mathcal{X}_{b,a}. \tag{1}$$

This fiber $\mathcal{F}_\mathbf{b}$ encapsulates the complete state vector $\boldsymbol{u}_b = (u_{b,1}, \dots, u_{b,|\mathcal{A}|})$ compatible with geometry $b$.

The total space $E$ of the bundle is then defined as the disjoint union of these fibers over the base manifold,

$$E = \bigsqcup_{b \in \mathcal{B}} \{b\} \times \mathcal{F}_b. \tag{2}$$

This structure is equipped with the canonical projection $\pi: E \to \mathcal{B}$ such that $\pi(b, u) = b$, which ensures that the inverse projection $\pi^{-1}(b)$ strictly recovers the valid function space $\mathcal{F}_b$ for the given geometry $b$.

Within this bundle structure, the physical dynamics are modeled as operations acting along the fibers. Because the temporal evolution of the system does not alter the geometric configuration, the state at time $t$ evolves to $t + \Delta t$ within the same fiber $\mathcal{F}_b$ for a fixed $b$. We define this evolution via a discrete operator family $\mathcal{S}_b: \mathcal{F}_b \to \mathcal{F}_b$ such that,

$$\boldsymbol{u}_b^{t+1} = \mathcal{S}_b(\boldsymbol{u}_b^t). \tag{3}$$

Collecting these operators across the entire base space yields a global bundle map $\boldsymbol{\Psi}$ acting on the total space,

$$\boldsymbol{\Psi}: E \to E, \boldsymbol{\Psi}(b, \boldsymbol{u}) = (b, \mathcal{S}_b(\boldsymbol{u})). \tag{4}$$

This formalism provides a natural interpretation of the dynamics as a bundle map acting along the fibers of the configuration dependent state space, capturing both the variability of battery geometries and the coupled evolution of heterogeneous physical fields in a single mathematical structure.

**Unified Fredholm Representation of the Multiphysics Operator Family**

The evolution map $\mathcal{S}_b: \mathcal{F}_b \to \mathcal{F}_b$ induced by the battery multiphysics equations lacks a tractable closed-form expression and offers no structural regularity to ensure consistency across varying geometries, domains, or equation types. To resolve this, we reformulate the governing coupled PDEs into a unified Fredholm integral framework. This formulation converts both transient and static PDEs into a canonical kernel-based operator family on each fiber, providing the mathematical foundation for the Strong Continuity proven in Theorem 1 of the Supplementary section.

Consider multiphysics state vector $\boldsymbol{u} \in \mathcal{F}_b$ on a specific geometry $b$. We define a generalized linear operator equation that encapsulates both transient and static physics,

$$\mathcal{M} \partial_t \boldsymbol{u}(t, x) + \mathcal{L}_b[\boldsymbol{u}](t, x) = \mathbf{q}(t, x), x \in \Omega_b. \tag{5}$$

Here, $\mathcal{L}_b$ denotes the spatial differential operator dependent on the domain topology, $\mathbf{q}(t, x)$ represents source terms. The capacity operator $\mathcal{M}$ serves as a selector for time-dependency. It entries are unity for transient fields such as temperature, and zero for static fields such as the electrolyte potential.

To derive the discrete time-evolution operator $\mathcal{S}_b$, we apply the Rothe method with a fixed time step $\Delta t$. Approximating the temporal derivative via an implicit Euler scheme yields the semi-discrete formulation,

$$(\mathcal{M} + \Delta t \mathcal{L}_b) \boldsymbol{u}^{t+1} = \mathcal{M} \boldsymbol{u}^t + \Delta t \mathbf{q}^t. \tag{6}$$

By defining the composite spatial operator $\mathcal{A}_b = \mathcal{M} + \Delta t \mathcal{L}_b$ and the aggregated source term $\mathbf{r}^t = \mathcal{M} \boldsymbol{u}^t + \Delta t \mathbf{q}^t$, the system reduces to a linear inversion problem at each time step $\mathcal{A}_b[\boldsymbol{u}^{t+1}](x) = \mathbf{r}_t(x)$. Assuming the boundary value problem is well-posed, the operator $\mathcal{A}_b$ admits a right inverse. From the theory of Green's functions, the solution $\boldsymbol{u}^{t+1}$ can be expressed explicitly as a Fredholm integral equation of the second kind,

$$\boldsymbol{u}^{t+1}(x) = \Phi_b[\mathbf{r}_t](x) = \int_{\Omega_b} \boldsymbol{K}(x, y; b)\, \mathbf{r}^t(y)\, dy. \tag{7}$$

Here, $\boldsymbol{K}(x, y; b)$ is the Green's function kernel matrix specific to the geometry $b$. Crucially, as the geometry $b$ varies across the base space $\mathcal{B}$, the associated Green's operator $\Phi_b$ changes accordingly.

For a multiphysics system with $N$ coupled fields, the kernel $\boldsymbol{K}$ is a block operator matrix,

$$\boldsymbol{K}(x, y; b) = \begin{pmatrix} K_{11} & \cdots & K_{1N} \\ \vdots & \ddots & \vdots \\ K_{N1} & \cdots & K_{NN} \end{pmatrix}. \tag{8}$$

In this matrix, a diagonal kernel $K_{ii}$ propagates the dynamics of the $i$-th physical field, while the off-diagonal kernel $K_{ij}(i \neq j)$ represents the cross-physics coupling influence, such as the effect of ohmic heat generation from the potential field on the temperature field. This block-Fredholm structure provides the theoretical blueprint for our neural architecture, which parametrizes the entire operator family $\mathfrak{F}$ by learning to approximate the geometry-dependent kernel matrix $\boldsymbol{K}(\cdot,\cdot; b)$ as a fiber-wise bundle map $\boldsymbol{\Psi}$.

**Architecture of Neural Bundle Map**

The architecture of the NBM is designed to computationally realize the bundle map $\Psi_\theta$ by structurally mimicking the Fredholm integral operation. To ensure consistency with the Strong Continuity requirement established in the Supplementary Information, the network facilitates the mapping of heterogeneous physical fields through a sequence of geometric embedding, field encoding, and multiphysics coupling.

For a given configuration $b$, the battery geometry is represented as a discretized point cloud $\boldsymbol{P}_b = \{p_k\}_{k=1}^{N_g} \subset \Omega_b$. A geometry encoder network $Geo_{enc}$ processes this spatial input to extract a global geometric latent descriptor $\mathbf{z}_g \in \mathbb{R}^{d_g}$, which is PointTransformer used in this work. To satisfy the topological axioms of the base manifold, $\mathbf{z}_g$ is projected onto a unit hypersphere $S^{d-1}$, serving as a locally compact Hausdorff base space. This Lipschitz-admissible embedding ensures that physical evolution varies continuously with respect to geometric perturbations, providing a stable foundation for the operator family.

Simultaneously, the input field for each physical field $\boldsymbol{u}_i$ is lifted into a feature space. Let $\{(\boldsymbol{y}_{m,i}, \boldsymbol{u}_i(\boldsymbol{y}_m))\}_{m=1}^M$ denote the set of coordinate-value pairs for the $i$ field. A field-specific encoder $\mathbf{MLP}_{enc,i}$ maps these coordinate-value pairs to local feature, which are then aggregated via a field-wise global pooling operation,

$$\mathbf{z}_{f,i} = \frac{1}{M} \sum_{m=1}^{M} \mathbf{MLP}_{enc,i}(\boldsymbol{y}_{m,i}, \boldsymbol{u}_i(\boldsymbol{y}_m)). \tag{9}$$

This pooling step serves as a Monte Carlo approximation of the integral operator $\int_{\Omega_b}(\cdot)dy$, generating a discretization-invariant representation $\mathbf{z}_{f,i}$ of the field state. In this refined architecture, the geometric descriptor $\mathbf{z}_g$ is concatenated directly with the field features $\mathbf{z}_{f,i}$ to produce a geometry-conditioned representation $\mathbf{h}_i$. This concatenation ensures that the physical dynamics are explicitly modulated by the boundary conditions and metric constraints encoded in $\mathbf{z}_g$.

The multiphysics coupling represented theoretically by the block kernel matrix $\boldsymbol{K}(\cdot,\cdot; b)$, is realized through a cross-field attention mechanism. To compute the updated state for the target field $i$, the network aggregates information from all source fields $j$.

$$\mathbf{c}_i = \sum_{j \in \mathcal{A}} \text{CrossAttention}(\mathbf{h}_i, \mathbf{h}_j). \tag{10}$$

The attention weights effectively learn the coupling strength between fields, such as the thermal influence of ohmic heat generation on the temperature field. This operation realizes the block-matrix multiplication inherent to the Fredholm representation of coupled PDEs.

The final decoding stage addresses the spatial heterogeneity of the fiber bundle. To reconstruct the solution $u_{t+1}^i$ for the $i$-th field, the coupled latent vector $\mathbf{c}_i$ is concatenated with query coordinates $x \in \Omega_{b,i}$ and processed by a field-specific decoder $\mathbf{MLP}_{dec,i}$.

$$u_i^{t+1}(x) = \mathbf{MLP}_{dec,i}(x, \mathbf{c}_i), \forall x \in \Omega_{b,i}. \tag{11}$$

This pointwise evaluation completes the neural approximation of the continuous operator field. As established in Theorem 2 of the Supplementary Information, this architecture possesses universal approximation capability for the class of geometry-dependent Fredholm operators defined on fiber bundles.

**Strong continuity and approximation of neural bundle maps**

The efficacy of the NBM rests upon two fundamental theoretical pillars: the existence of a continuous evolution operator across varying geometries and the ability of the proposed architecture to approximate this operator with arbitrary precision.

**Strong continuity of the multiphysics operator family.** We first establish that the variation of battery physics with respect to geometry is mathematically well-posed. Let $\mathcal{H}$ be the continuous field of Hilbert spaces over the geometric base space $\mathcal{B}$, where each fiber $\mathcal{H}_b = L^2(\Omega_b)$ represents the state space for a specific cell design. The temporal evolution of the physical fields is governed by operator family $\mathcal{S} = \{S_b: \mathcal{H}_b \to \mathcal{H}_b\}_{b \in \mathcal{B}}$.

Drawing upon the theory of operator fields over varying domains, the space of bounded operators $\mathcal{L}^s(\mathcal{H})$ equipped with the strong operator topology forms a continuous field of locally convex spaces. Since our geometric embedding maps design parameters to a latent manifold, i.e., the hypersphere, the base space $\mathcal{B}$ is locally compact Hausdorff space. Under these conditions, the existence of continuous local frames guarantees that if two geometries $b_1, b_2$ are close on the manifold, the associated physical operators $S_{b_1}$ and $S_{b_2}$ are close in the operator fields. This strong continuity, defined as $b \mapsto S_b \in \mathcal{L}^s(\mathcal{H})$ is the prerequisite for learnability, ensuring that the target function for our neural network does not exhibit chaotic discontinuities under geometric perturbation.

**Universal approximation via neural kernels.** Having established the continuity of the target operator family, we demonstrate that the NBM architecture serves as a universal approximator for continuous operator fields. Our architecture approximates this geometry-dependent kernel function $K(\cdot,\cdot;b)$ through the integration of the latent geometric variable $\mathbf{z}_g$ and the cross-field attention mechanism.

By extending the Universal Approximation Theorem for operators to fiber bundles, we state in Theorem 2 that for any $\epsilon > 0$, there exist and NBM configuration such that the neural operator $\mathcal{N}$ satisfies,

$$\sup_{b \in B} \| S_b - \mathcal{N}_b \|_{\mathcal{L}(H_b)} < \epsilon. \tag{12}$$

This confirms that the NBM is not merely a heuristic surrogate, but a rigorous discretization of the continuous operator field governing multiphysics evolution. (Detailed proof is provided in Supplementary Information Section 1).

**Acknowledgements**

This research was funded by the National Natural Science Foundation of China under Grant 52575579 (Changqing Liu), UK EPSRC Grant EP/V007335/1 (Yan Jin), Royal Society Research Grant RGS\R2\252634 (Jie Lin), UKRI EPSRC DICE Networks+ Flexible Fund grant (Jie Lin) and Royal Society International Exchanges Program IEC\NSFC\233361 (Jie Lin).

**Author Contributions Statement**

Y.L. supervised the project and obtained funding for the research. Z.Z., C.L., J.L., and Y.L. conceptualized the study. Z.Z. and C.L. performed formal analysis and contributed the original draft. Z.Z., C.L., J.L., F.Y, Y.Z., Y.J. and Y.L. contributed to writing, review, and editing. All authors have read and agreed to the final published version of the manuscript.

**Competing interests and ethics declaration**

The authors declare no competing interest.